\documentclass{PoS}

\title{
\vspace*{-2cm}
{\normalsize {\rm \hfill\textnormal{LLNL-PROC-679800} }} \\
\vspace*{1cm}
S-parameter and vector decay constant in QCD with eight fundamental
fermions}

\ShortTitle{S-parameter and vector decay constant in $N_f=8$ QCD}

\author{
  \speaker{Yasumichi Aoki}$^a$,
  Tatsumi Aoyama$^a$,
  Ed Bennett$^{b,a}$,
  Masafumi Kurachi$^c$,
  Toshihide~Maskawa$^a$,
  Kohtaroh Miura$^{d,a}$,
  Kei-ichi Nagai$^a$,
  Hiroshi Ohki$^e$, 
  Enrico~Rinaldi$^f$,
  Akihiro Shibata$^g$,
  Koichi Yamawaki$^a$
  and 
  Takeshi Yamazaki$^h$ 
  
  \hspace*{50mm} (LatKMI Collaboration) 
  \\
  
  $^a$
  Kobayashi-Maskawa Institute for the Origin of Particles and the Universe (KMI), Nagoya University, Nagoya 464-8602, Japan \\
  $^b$
  Department of Physics, Swansea University, Singleton Park, Swansea SA2 8PP, UK \\
  $^c$
  Institute of Particle and Nuclear studies, High Energy Accelerator
  Research Organization (KEK), Tsukuba 305-0801, Japan \\
  $^d$
  Centre de Physique Theorique(CPT), Aix-Marseille University, Campus de
  Luminy, Case 907, 163 Avenue de Luminy, 13288 Marseille cedex 9,
  France \\
  $^e$
  RIKEN BNL Research Center, Brookhaven National Laboratory, Upton, NY
  11973, USA \\
  $^f$ 
  Nuclear and Chemical Sciences Division, Lawrence Livermore National Laboratory, Livermore, CA 94550, USA \\
  $^g$
 Computing Research Center, High Energy Accelerator Research
  Organization (KEK), Tsukuba 305-0801, Japan \\
  $^h$
  Graduate School of Pure and Applied Sciences, University of Tsukuba, Tsukuba, Ibaraki 305-8571, Japan\\
}

\abstract{
SU(3) gauge theory with eight massless fundamental fermions seems to be
near the conformal boundary, and is a candidate theory of walking
technicolor. Along the series of study by LatKMI   
collaboration using HISQ fermions, $S$-parameter and vector decay
constant, which provide important constraints in the model of
electroweak symmetry breaking, are calculated for this theory. 
Use of various volumes allows a systematic investigation of 
finite volume effects. A strong sensitivity of the $S$-parameter to the
volume is found.
}

\FullConference{The 33rd International Symposium on Lattice Field Theory\\
                 14 -18 July  2015\\
                 Kobe International Conference Center, Kobe, Japan}

\begin{document}

\section{Introduction}

Many-flavor QCD provides a possibility for the strong dynamics 
realization of the Higgs sector in the Standard Model.
Among the tested models, SU(3) gauge theory with eight mass-less
fundamental fermions draw much attention recently\footnote{
see for example reviews by Hasenfratz (these proceedings) and
Kuti \cite{Kuti:2014epa}.}, 
especially as it could be a successful candidate of walking technicolor 
theory with a flavor-singlet scalar possibly light enough to realize 
125 GeV Higgs \cite{Aoki:2013qxaF}.
In addition to the tests of walking dynamics and light Higgs realization,
the Peskin-Takeuchi $S$ parameter 
\cite{Peskin:1991sw}
provides an important constraint.
As increasing number of fermions would naturally enhance $S$,
there is a concern that 
many-flavor models will critically face the limit set by electroweak precision
measurements. 

The $S$ parameter in the many flavor theories has been studied 
first by using Bethe-Salpeter equation \cite{Harada:2005ru,Kurachi:2006mu}.
Lattice computation of the $S$ parameter is first performed using the
overlap fermions \cite{Shintani:2008qe} and domain-wall fermions
\cite{Boyle:2009xi} on the conventional QCD configurations,
mainly aiming to test a methodology. 
Chiral symmetry at finite lattice spacing is essential for the
computation, which is the reason why the chiral fermion formulation is
used in these studies.

In the context of strong dynamics theory for the electroweak symmetry
breaking, LSD collaboration is the first to systematically study
the $S$ parameter on the dependence to the number of flavors $N_f$ adopting the 
domain-wall-fermion formulation in lattice gauge computation.
The results are reported for SU(3) gauge theory with 
$N_f=2$, 6 \cite{Appelquist:2010xv} and 8 fundamental fermions
\cite{Appelquist:2014zsa}.
LSD results indicate a decreasing trend of $S$ towards
lighter fermion mass $m_f$ for $N_f=6$ and 8, which would ease 
the model building with these theories. The trend is also backed up by the
tendency of the observed spectrum towards parity doubling. One caveat,
if any, that these studies have is that only one spatial volume is used
thus the finite volume effects 
might have been overlooked. This is particularly important in this context
as the finite volume effect tends to reduce the degree of chiral
symmetry breaking, which could drive the system towards parity doubling,
hence reduce $S$.

We have been studying $N_f=8$ SU(3) gauge theory 
with hadron spectrum as a main probe, where the indication of the theory
being close to the 
conformal phase boundary and a parametrically light flavor singlet scalar 
are found \cite{Aoki:2013xzaF,Aoki:2013qxaF,Ohki:lat2015}.
Therefore, we regard $N_f=8$ SU(3) gauge theory as a candidate of
walking technicolor theory, which has a potential to coop with the
experiments. 
We use the highly-improved staggered quarks (HISQ) throughout our 
many flavor studies for $N_f=4$, 8, 12, 16. To fully utilize the 
gauge configuration ensembles so far accumulated to a good statistics,
staggered-fermion calculation for the $S$ parameter is required.
We developed the formulation and reported initial results
in the past lattice conferences\footnote{Lattice 2013 talk by Y.~Aoki
and Lattice 2014 talk by H.~Ohki.}. Here we report on our computation of
the $S$ parameter for $N_f=8$ and provide an update.

The calculation method, which makes use of the exact chiral symmetry
on staggered fermions with multiple staggered field,
is briefly described and the preliminary results will be 
reported in the next section. As a byproduct of the measurements of
$S$ parameter, the vector meson decay constant is also measured and
reported.

\section{Vacuum Polarization Functions and Spectrum}

Peskin-Takeuchi $S$ parameter is calculated from the slope of
the transverse $V-A$ vacuum polarization function $\Pi(q^2)$ with respect
to the infinitesimal squared momentum transfer $q^2$,
\begin{eqnarray}
 S & = & 4\pi \Pi'(q^2=0), \label{eq:S}\\
 \Pi_{\mu\nu}(q) & = & \delta_{\mu\nu}\Pi(q^2) 
  -\frac{{q}_\mu{q}_\nu}{{q}^2}\tilde{\Pi}(q^2),\\
 \Pi_{\mu\nu}(q) & \equiv & \int dx^4 e^{iqx}
  \left\{ \langle V^{(1,2)}_\mu(x) V^{(2,1)}_\nu(0)\rangle
   - \langle A^{(1,2)}_\mu(x) A^{(2,1)}_\nu(0)\rangle \right\},\\
\end{eqnarray}
where $V^{(i,j)}_\mu$ and $A^{(i,j)}_\mu$ are the flavor non-singlet
vector and axial vector current respectively, made of anti-quark
field with $i$-flavor and quark field with $j$-flavor.
This defines the contribution of $S$ parameter from one electro-weak doublet.

On the lattice a naive definition of $S$ parameter may easily
suffer from ultra-violet power-divergent contribution due to a lack of
continuum symmetry\cite{Shintani:2008qe}. 
An exact symmetry which transforms $V_\mu$ into $A_\mu$ and vice versa
can be used to cancel the UV divergence in the difference.
In the literature overlap fermions or domain-wall fermions are used for
the computation of $S$ parameter
\cite{Shintani:2008qe,Boyle:2009xi,Appelquist:2010xv,Appelquist:2014zsa}
due to this reason.
However such calculations are computationally expensive.

As we use staggered fermions, namely HISQ, it is  useful to have
a formulation to calculate $S$ parameter within HISQ. Staggered fermions
have a exact flavor and chiral symmetry which is part of the full flavor
and chiral symmetry in the continuum. This flavor symmetry is of the
baryon number. The associated current is the flavor singlet vector 
current in four-flavor (one-species) theory, thus is not suitable
to compute the $S$ parameter. Fortunately our primary interest is for
the eight flavors, where the exact symmetry enhances to contain
$SU(2)_L\times SU(2)_R$. Therefore the flavor non-singlet
non-renormalized vector and axialvector currents can be constructed
as the Noether currents of the symmetry.

The conserved currents of HISQ involve length-one and three point-split
bilinears. All have the same spin-taste structures
\begin{equation}
 \begin{array}{cc} 
 (\gamma_\mu\otimes 1)\tau^a, & (\gamma_\mu\gamma_5\otimes\xi_5)\tau^a\\
 \end{array}
\end{equation}
for vector and axial vector respectively. $\tau^a$, being $SU(2)$ generators
for the eight-flavor theory, bind two different staggered fields.
The length-three operators are from the Naik term in the action.
Apart from the spin-taste structure, the vector and the axial vector
currents are exactly same. The exact chiral symmetry with the generator
$(\gamma_5\otimes\xi_5)\tau^a$ transforms one into another.
It is this transformation  property that ensures the cancellation of the
UV power divergence when the $V-A$ difference of the vacuum polarization
function is composed. Therefore, the use of the conserved current is not
necessary in general. For example, the operator with just the length-one
part from the conserved current will do the job, though renormalization of the
operator is then necessary.

In practice we chose to use the one-link point-split currents at the
source position of the two-point function and the full conserved
currents at the sink. This makes both the numerical effort and coding
complexity minimal while keeping useful Ward-Takahashi identities intact.
The necessary renormalization factor for the one-link operator is
calculated by taking the ratio:
\begin{equation}
 Z_A(t) = \frac{\langle {\mathcal A}_4(t) P(0)\rangle}
{\langle A_4(t) P(0)\rangle},
\end{equation}
where ${\mathcal A}_4(t)$ and $A_4(t)$ are the temporal component 
of the conserved and one-link axial vector current respectively with
zero-momentum projection at time $t$. $P(0)$ is the pseudoscalar
operator with the structure $(\gamma_5\otimes\xi_5)$ at the origin. 
The axial current renormalization $Z_A$ for ${\mathcal A}_\mu=Z_A A_\mu$
is determined at plateau at large $t$.
\begin{figure}[tbh]
\begin{center}
 \includegraphics*[angle=0,width=0.45\textwidth]{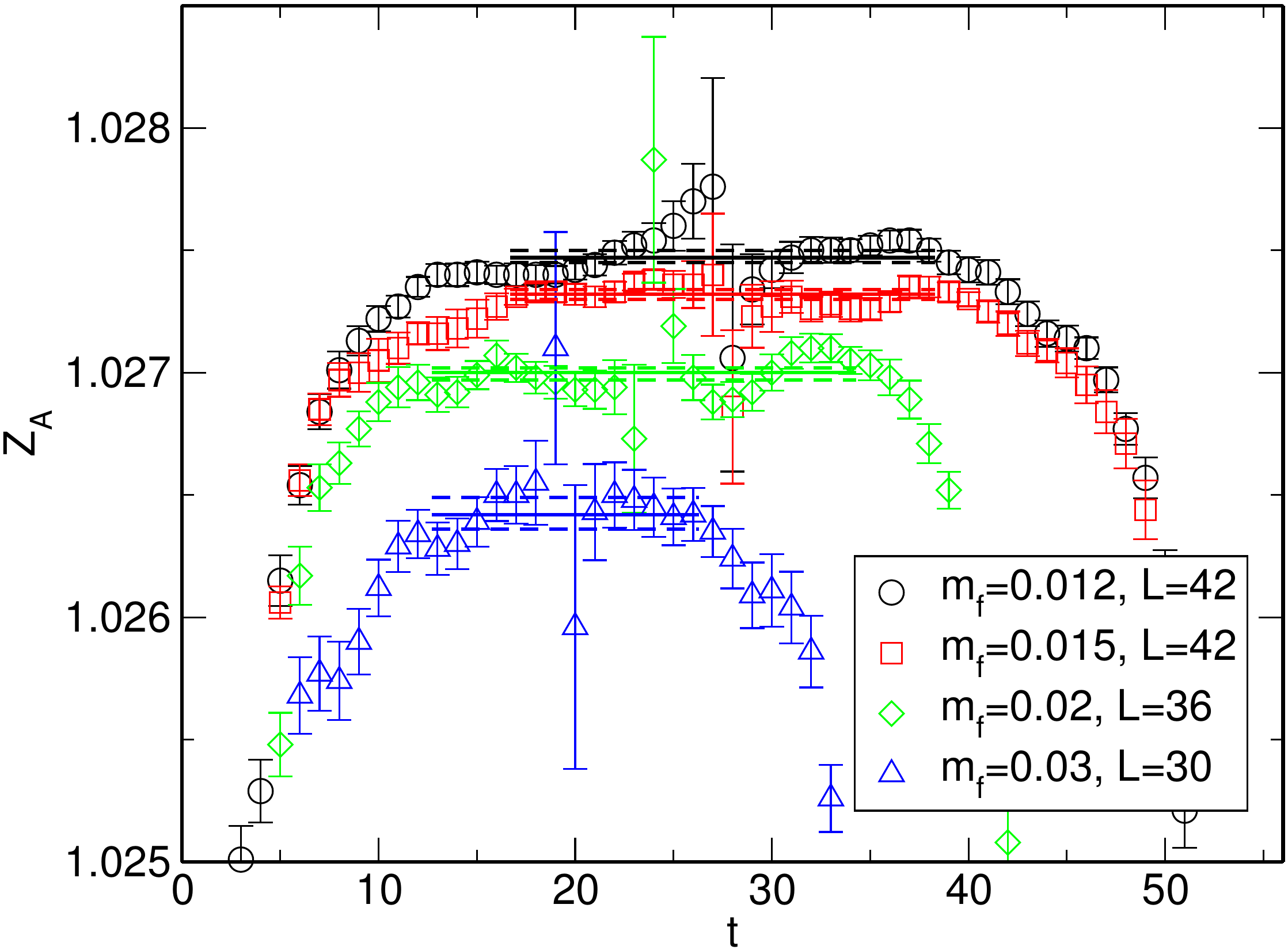}
 \ 
 \includegraphics*[angle=0,width=0.45\textwidth]{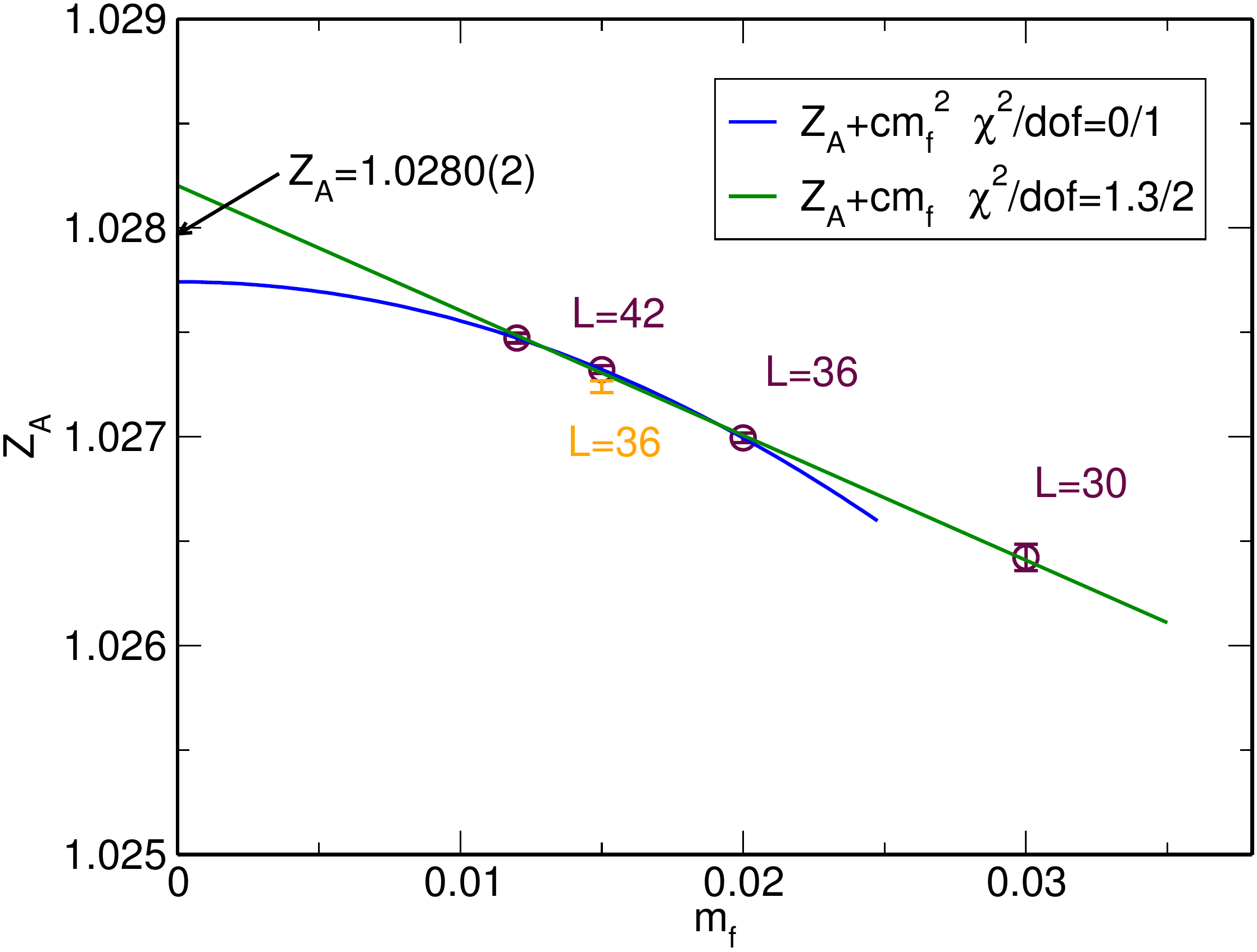}
\end{center}
\caption{Ratio and plateau fitting for $Z_A$ (left) and chiral
 extrapolation with linear or quadratic in $m_f$ (right).}
\label{fig:ZA}
\end{figure}
The left panel of Figure \ref{fig:ZA} shows the ratio $Z_A(t)$ for
the so-far smallest masses. 
Right panel shows the extrapolation to the
chiral limit using these four points, where we use the largest volume
for each mass $L=30$ ($m_f=0.03$), 36 ($m_f=0.02$) and 42 ($m_f=0.015$
and 0.0125) for $L^3\times T$ ($T=4/3*L$) lattice.
 Both linear extrapolation with all four
points and quadratic with lightest three are good. We adopt the average
for our $Z_A$ and the systematic error of the extrapolation is estimated
from the difference from the average to linear or quadratic, which is small
(0.02\%). The statistical error is less than 0.005\%,
thus negligible.

For the $S$-parameter, vacuum polarization form the vector current
two-point function is needed, too. The one-link vector current is
renormalized with $Z_A$ so far obtained by taking advantage of the exact
chiral symmetry, which leads to $Z_V=Z_A$.

\begin{figure}[tbh]
\begin{center}
 \includegraphics*[angle=0,width=0.45\textwidth]{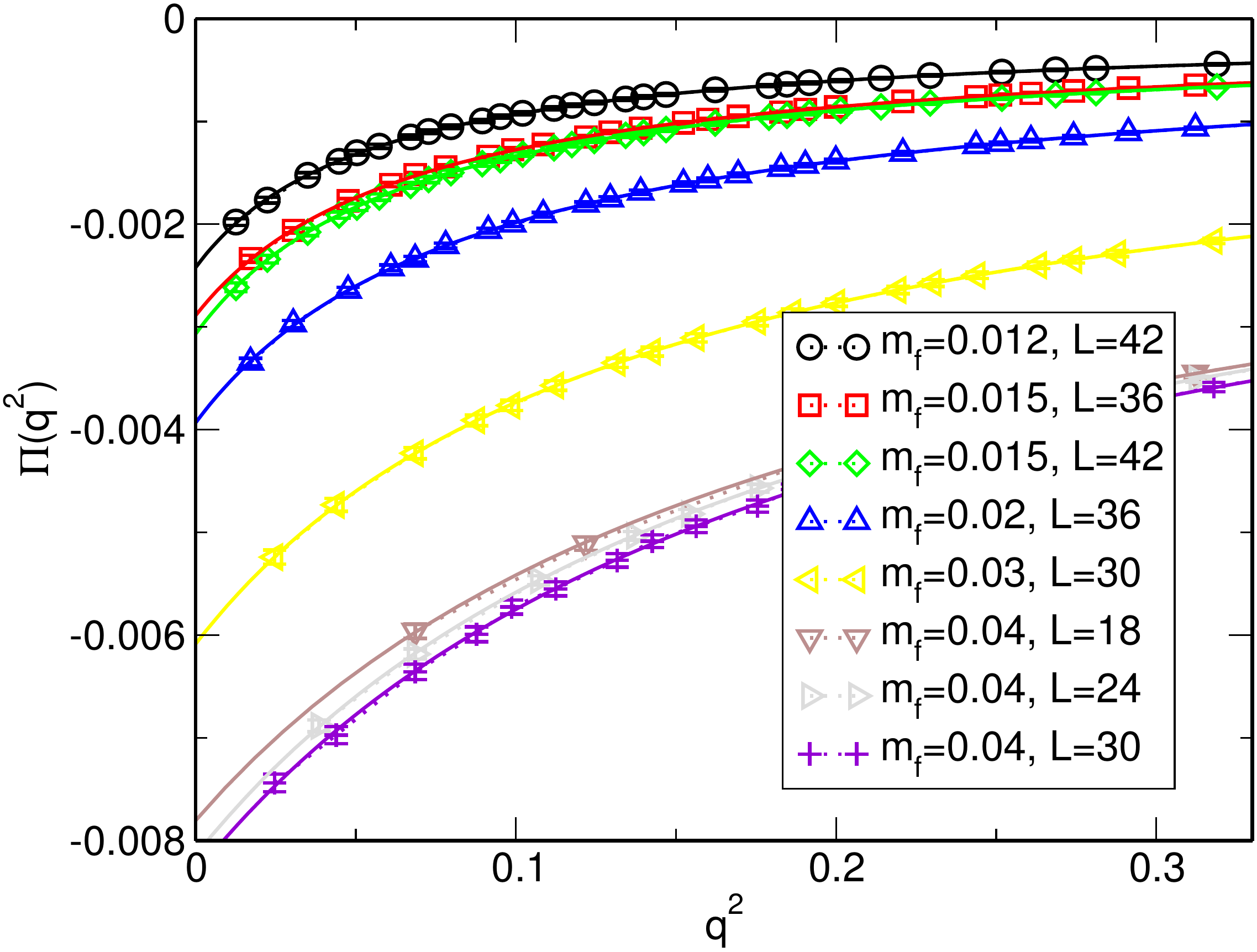}
\end{center}
\caption{Bare $V-A$ transverse vacuum pluralization function $\Pi(q^2)$. Solid lines show fits with the Pad\'e approximation
 $f(q^2) = (b_0 + b_1 q^2)/(1 + c_1 q^2 + c_2 q^4)$.}
\label{fig:Pi-q2_fit}
\end{figure}

The results of $V-A$ transverse vacuum polarization function $\Pi(q^2)$
are shown in Fig.~\ref{fig:Pi-q2_fit}. These results are from the bare
two point functions. The one-link operator at the source position
needs to be renormalized, which will be taken care of by multiplying one
factor of $Z_A$ to the results. The Solid lines show fits with the
Pad\'e approximation 
 $f(q^2) = (b_0 + b_1 q^2)/(1 + c_1 q^2 + c_2 q^4)$ \cite{Appelquist:2010xv}.

\begin{figure}[tbh]
\begin{center}
 \includegraphics*[angle=0,width=0.44\textwidth]{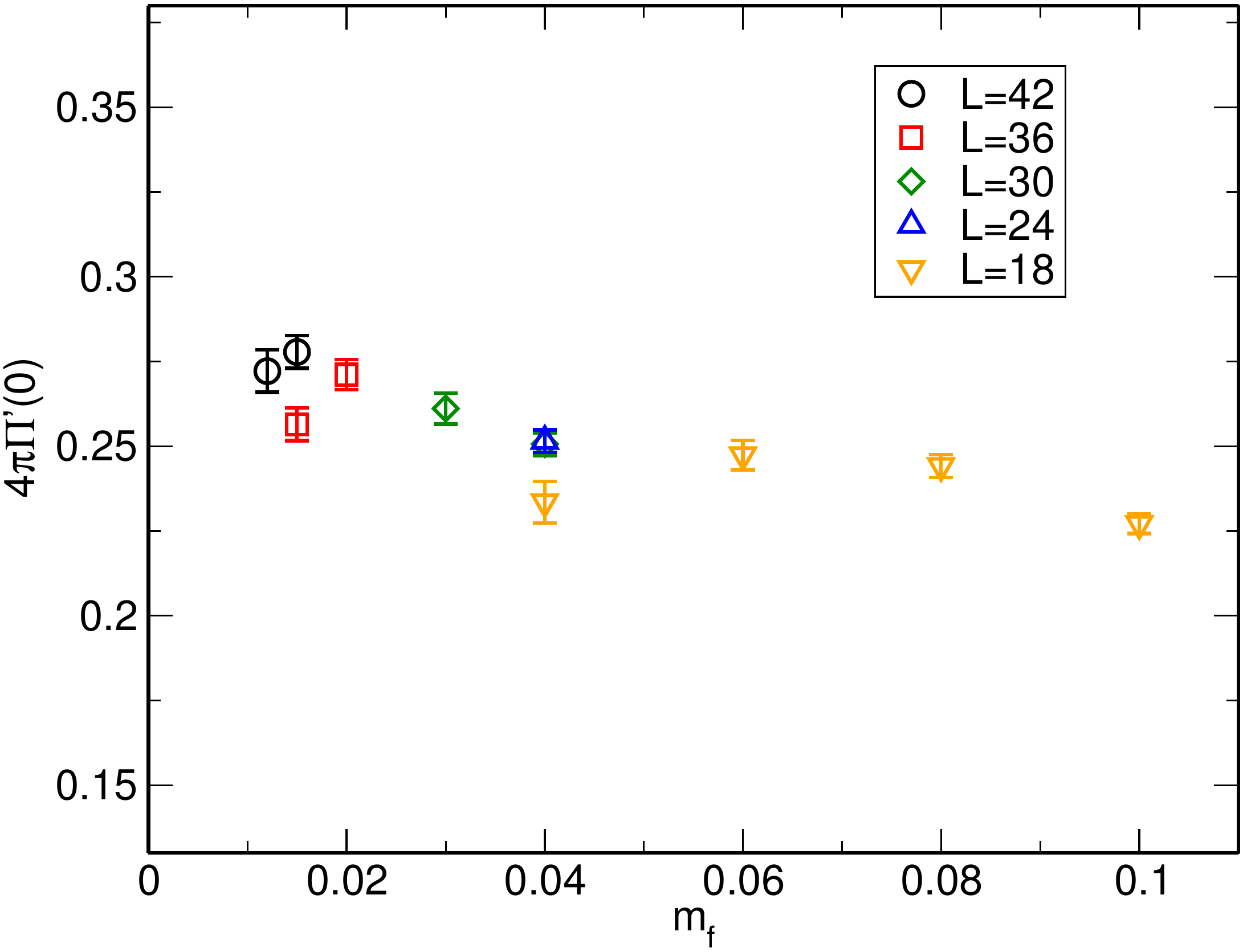}
 \hspace{12pt}
 \includegraphics*[angle=0,width=0.45\textwidth]{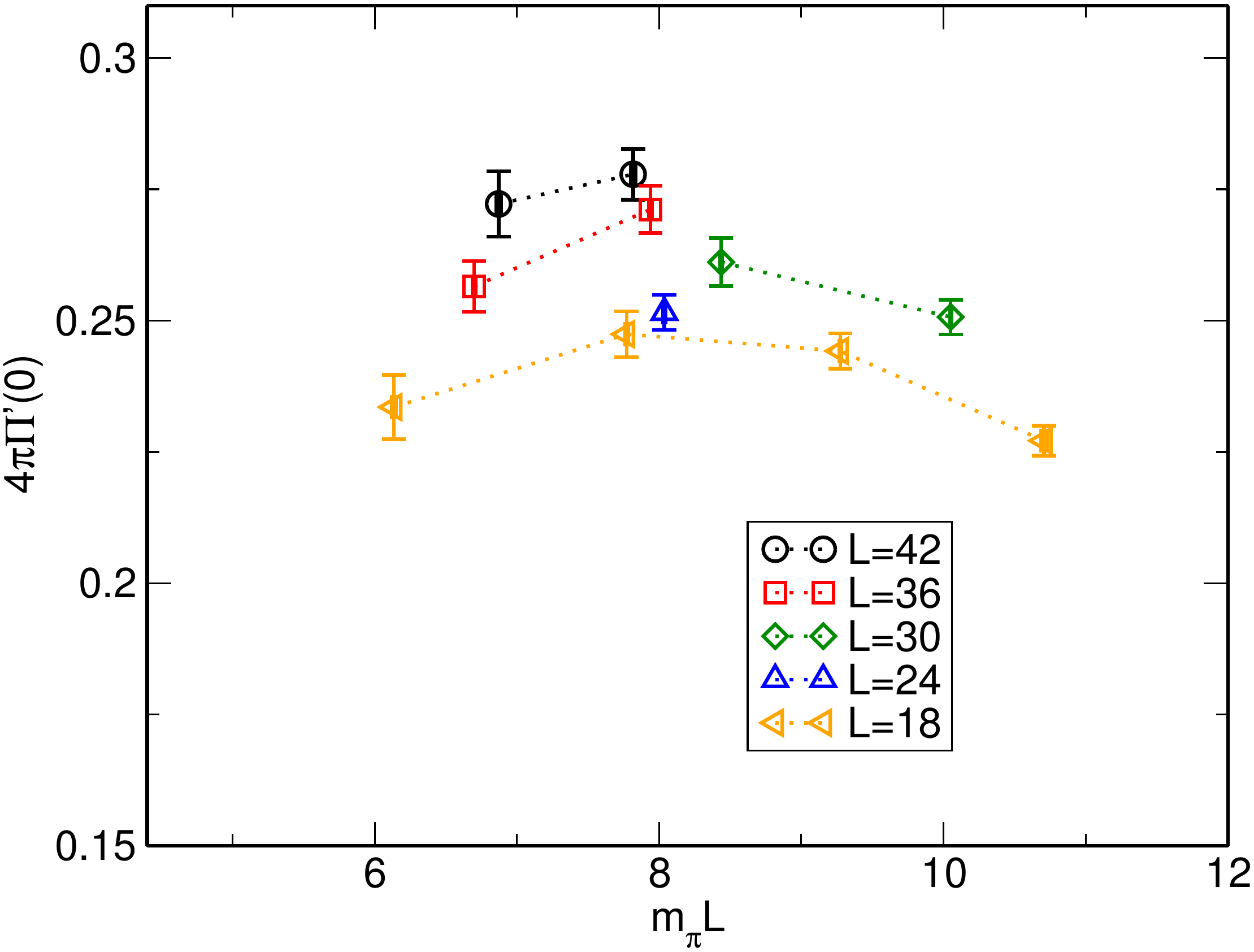}
\end{center}
 \caption{Strong-dynamics contribution to the $S$ parameter per
 electroweak doublet as a function of quark mass $m_f$ (left) or
 of a dimension-less product of techni-pion mass $m_\pi$ and the linear
 spatial size $L$ (right). }
\label{fig:S}
\end{figure}

The strong-dynamics contribution to the $S$ parameter per
electroweak doublet, Eq.~(\ref{eq:S}) is
calculated through the Pad\'e fit and shown in Fig.~\ref{fig:S} 
as a function of quark mass $m_f$ (left panel).
Multiplication of $Z_A$ has been done to get renormalized $S$ parameter.
Following the same color toward smaller $m_f$, $S$
bends down. This movement is of a finite volume effect as changing to
larger volume will move the point back upward. The size of the effect
is quite large. For example, the change of $S$ at $m_f=0.015$ from
$L=36$ to $42$ is 8\%, while the pion mass shift at the same point
is only 0.04\% and statistically consistent with zero.
Note that for the test against experiment, subtraction of the standard
Higgs sector from this result is needed
\cite{Peskin:1991sw,Schaich:2011qz}. The finite volume effect
discussed here, however, is independent from that.

The right panel drawn against $m_\pi L$, a dimension-less
product of techni-pion mass $m_\pi$ and the linear spatial size $L$,
 shows a same structure for 
different $L$. The left-most points of $L=18$ and $36$ suffer from
the finite volume effect, which is apparent from the left panel.
The bending-down seems to occur around $m_\pi L\simeq 7$. As the smallest 
mass point $m_f=0.012$ with $L=42$ is in the range of $m_\pi L< 7$,
the finite volume effect is suspected.

\begin{figure}[tbh]
\begin{center}
 \includegraphics*[angle=0,width=0.45\textwidth]{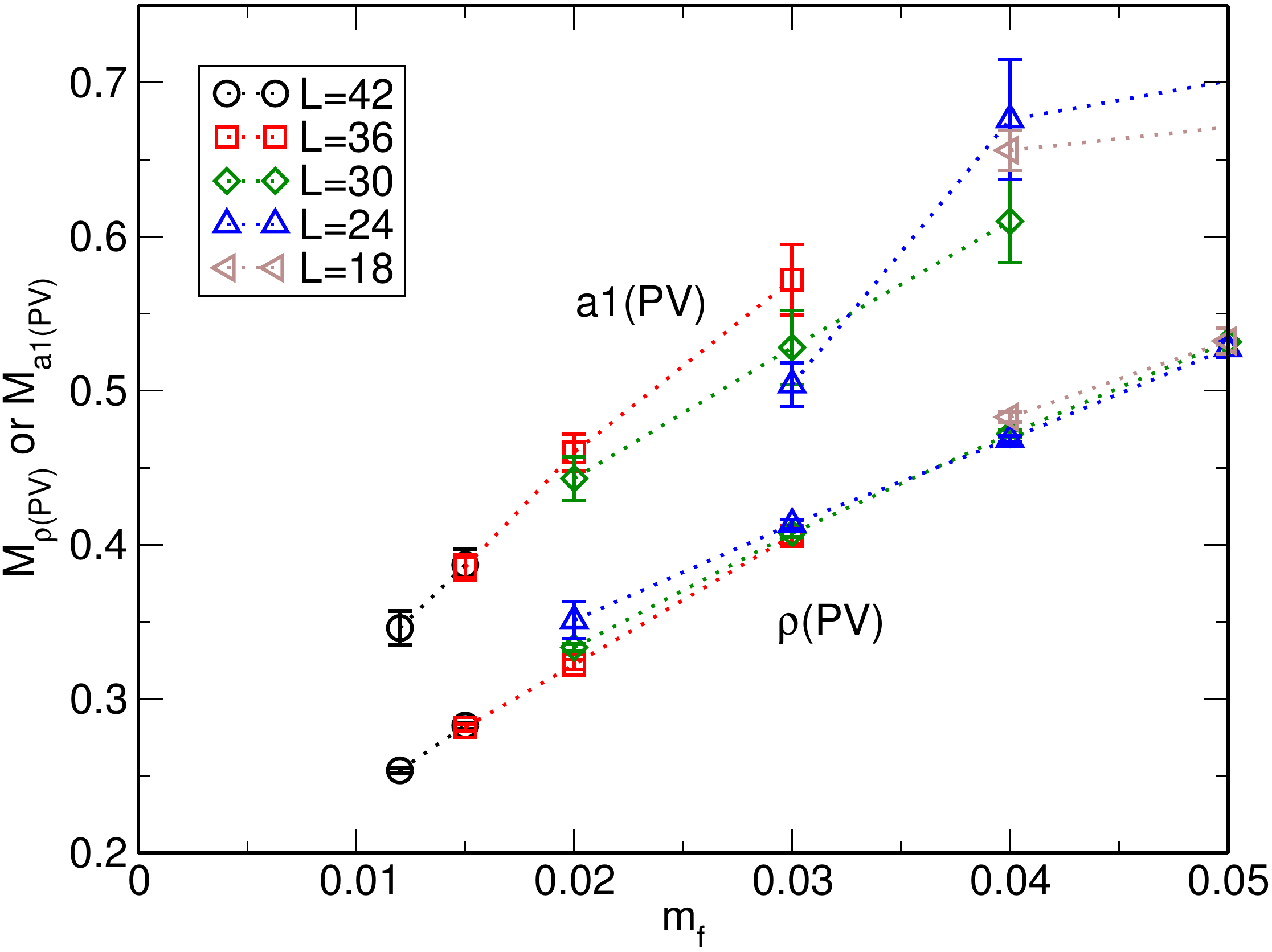}
 \hspace{12pt}
 \includegraphics*[angle=0,width=0.45\textwidth]{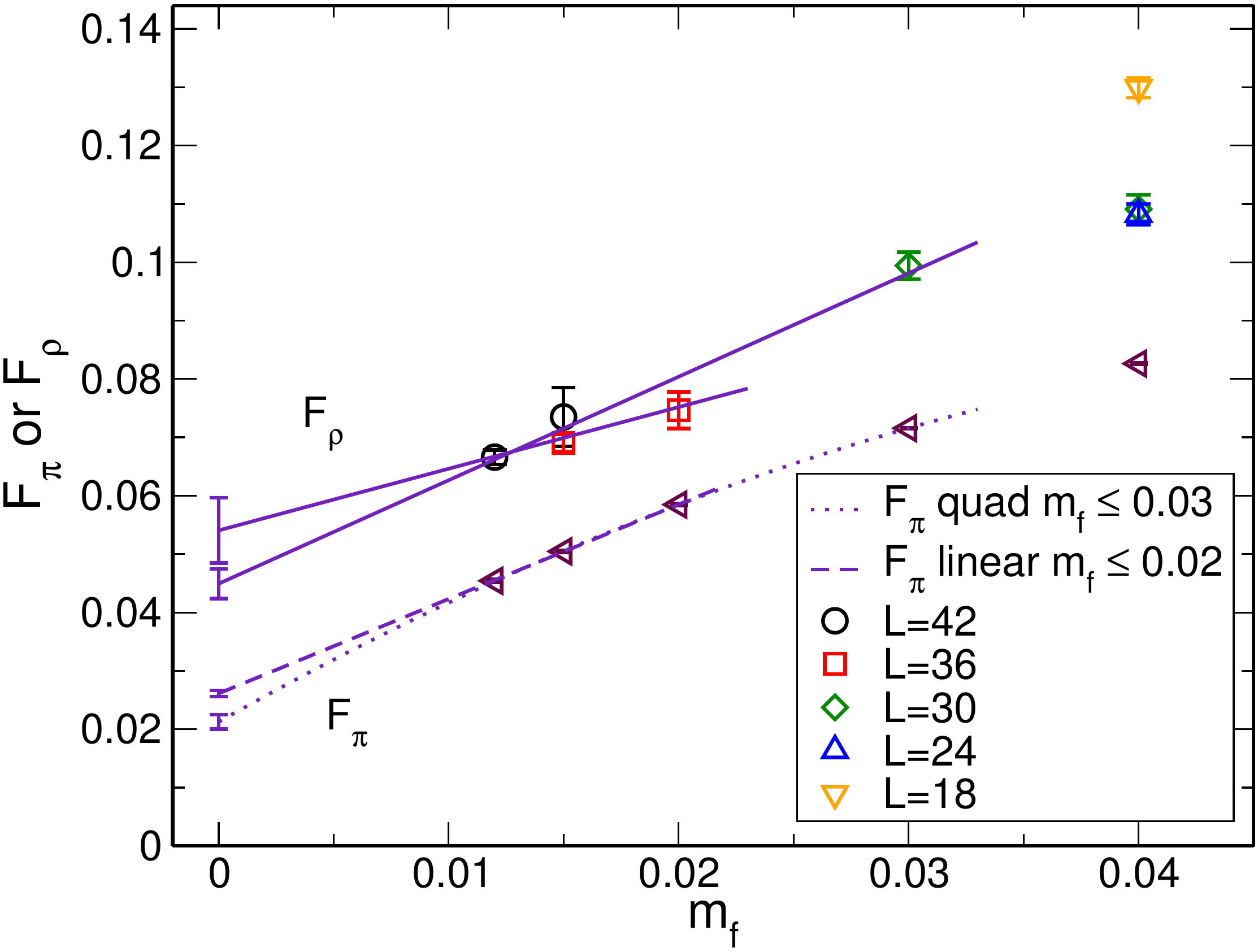}
\end{center}
 \caption{Vector ($\rho$) and axialvector ($a_1$) meson masses measured 
 with $PV$ channel as a function of quark mass (left). Vector meson
 decay constant through the smeared one-link current is plotted (colored
 symbols) in comparison with the pion decay constant (black symbols)
 (right). Linear extrapolation for the chiral limit $F_\rho$ is
 performed with two different mass ranges.
}
\label{fig:spectrum}
\end{figure}
The left panel of Fig.~\ref{fig:spectrum} shows the vector and
axialvector meson masses  measured with the PV channel
$(\gamma_k\gamma_4\otimes \xi_k\xi_4)$. 
Although this channel is different from the one that is used to calculate $S$,
we use this as the signal is much better than that due to the locality
of the operator in the staggered hypercube. We expect that good taste
symmetry, which has been confirmed in the pion sector, also holds for
the these sectors. 

A remarkable finite volume dependence is observed for the vector meson
mass, which drives the meson heavier as the volume is reduced.
Due to the larger statistical noise the situation is not very clear for
the axialvector. However the for $m_f=0.02$ and 0.03, the effect appear
opposite to the vector meson. The movement to the small volume is
towards the parity doubling, thus seems consistent with the volume
effect on $S$ through the dispersive analysis using spectral
decomposition \cite{Peskin:1991sw}.
We have to note, however, the correspondence between the parity doubling and
the decrease of $S$ is not entirely clear with the current result. For example,
as already noted $S$ decreased 8\% from $L=42$ to 36 at $m_f=0.015$,
while the meson masses at the same parameter are stable.

In the spectral decomposition of the $S$ parameter
\cite{Peskin:1991sw}  not only the mass but also the decay
constant enters. 
The right panel of Fig.~\ref{fig:spectrum} show the vector meson 
decay constant computed through the symmetric two-point function with
the one-link operator in both source and sink position.
There, significant volume dependence is not observed except for the
smallest volume, $L=18$.

One also needs to study the axialvector decay constant, which is not
shown in this report and  remain to be a future task. 
With our data the vector and  axialvector decay
constant can in principle be determined. However the axial vector channel
is an excited state in the two point correlation function, therefore quite
noisy. This is, unfortunately, a drawback of the use of staggered fermions. 

Let us get back to the vector decay constant and try to infer the value 
in the chiral limit. A rough analysis leads to the chiral limit
value of the decay constant which appear approximately twice as the pion 
decay constant $f_\rho \simeq 2 f_\pi$. This is consistent with the results
obtained by LSD collaboration \cite{Appelquist:2014zsa} with domain-wall
fermions.

\section{Summary and Outlook}

In this report we briefly described how the Peskin-Takeuchi $S$ parameter
is computed using Highly Improved Staggered Quarks (HISQ) and 
preliminary results obtained by applying the method to the eight-flavor
SU(3) gauge theory were shown. A remarkable strong sensitivity of 
the $S$ parameter to the system volume was found, which indicates a careful
volume study is indispensable to obtain the correct chiral behavior of the $S$
parameter. The spectrum closely related with $S$ parameter includes 
vector and axialvector meson masses and decay constants. Noise reduction,
where a brute-force method might suffice, especially on the axialvector 
sector will be useful for an in-depth understanding of the finite volume
effect on $S$.  
$V-A$ vacuum polarization can be used as a probe of (near)
conformality through the mass scaling \cite{DelDebbio:2010jy}, 
which will be of interest when the finite volume issues are settled.

\vspace{12pt}

{\it Acknowledgments} --
Numerical computations have been carried out on $\varphi$ at KMI, CX400
at the Information Technology Center in Nagoya University, and CX400 and
HA8000 at the Research Institute for Information Technology in Kyushu
University. This work is supported by the JSPS Grant-in-Aid for
Scientific Research (S) No.22224003, (C) No.23540300 (K.Y.), for Young
Scientists (B) No.25800139 (H.O.) and No.25800138 (T.Y.), and also by
the MEXT Grants-in-Aid for Scientific Research on Innovative Areas
No.23105708 (T.Y.) and No.25105011 (M.K.). This work is supported by the
JLDG constructed over the SINET of NII. The work of H.O. is supported by
the RIKEN Special Postdoctoral Researcher program. E.R. acknowledges the
support of the U.S. Department of Energy under Contract 
DE-AC52-07NA27344 (LLNL). 

\bibliography{paper_yaoki}

\end{document}